\documentclass[%
preprint,
 amsmath,amssymb,
 aps,
]{revtex4-2}

\usepackage{graphicx}
\usepackage{dcolumn}
\usepackage{bm}

\begin{document}


\title{Surface energy-driven crumpling transition in a thin sheet under compression}
\author{Aashna Chawla}
\author{Deepak Kumar}
 \email{krdeepak@physics.iitd.ac.in}
\affiliation{
 Department of Physics, Indian Institute of Technology Delhi, Hauz Khas, New Delhi-110016, India.}
 
\begin{abstract}
In our common experience, crumpling a sheet requires external compressive force and leads to a random network of folds. However, thin sheets have been theoretically predicted to spontaneously transition from a flat to a crumpled state driven by thermal fluctuations, a phenomenon that has been elusive in experiments. We report the first observation of a similar crumpling transition driven instead by surface energy. Using a sensitive experimental protocol, when we gently compress a thin polymer sheet weakly adhered to a hydrogel substrate it transitions to a self-crumpling state at a well-defined critical compression independent of system details. The transition is marked by the percolation of a fold network, and a power-law increase in fold density. Most remarkably, the crumpled state shows a tunable order of folds establishing the phenomenon’s potential as a simple and scalable technique to do origami with extremely thin sheets. 
\end{abstract}

\maketitle

\clearpage

\section*{Introduction}
Thin sheets having a much larger lateral size than thickness are an extremely common material geometry encountered in nature, in our day-to-day lives, in research labs, and in industries. Due to their ease of bending, these materials readily deform out of their intrinsic shape, bearing an intricate, and often beautiful, pattern of buckled structures. Examples ranging in scale from geological to the nanoscopic abound: mountainous relief formed in the earth’s crust due to tectonic activity, wavy patterns in hanging curtains\cite{vandeparre2011wrinkling}, wrinkles in polymeric sheets floating on liquid surfaces\cite{huang2007capillary}, wrinkling of skin due to ageing\cite{cerda2003geometry}, and unavoidable wrinkles sustained in sheets of graphene during device fabrication processes\cite{deng2016wrinkled}. In particular, due to the revolutionary impact of 2D materials such as graphene and $\mathrm{MoS_2}$ on the world of nanotechnology, there is an imperative to comprehensively understand the physics of large  deformations seen in extremely thin sheets. 

A useful example of such large deformation in thin sheets is presented by origami wherein by folding a thin flat sheet in an ordered pattern one can not only carve it into various desired 3D structures but also build non-trivial mechanical response and functionality such as auxeticity \cite{cohen_2014}. Another example, part of even more common experience, is offered by a randomly crumpled piece of paper. In contrast to origami, crumpling is characterized by a dense disordered network of folds but is associated with equally remarkable and rich physical phenomenology, including renormalized bending and stretching stiffnesses\cite{kovsmrlj2013mechanical}, slow relaxation\cite{matan2002crumpling}, and aging and memory effects\cite{shohat2022memory}. Even though the fundamental mechanism underlying these processes are known to be scale-invariant and expected to work as well for even extremely thin sheets such as 2D electronic materials like graphene\cite{blees2015graphene}, the lack of suitable experimental techniques for imprinting a desired pattern of creases or folds in such thin sheets has limited their application at the nano-scale. 

The compressive force required to produce such large deformations is normally expected to increase with compression\cite{matan2002crumpling}. Moreover, when the compressive force is removed, the sheet usually relaxes toward the uncrumpled configuration. In contrast, an opposite type of effect is predicted for thermalized thin sheets, where above a critical temperature, the sheet spontaneously transitions into a crumpled state\cite{kantor1987crumpling, Abraham1991}. This entropy-driven transition is expected to occur at a temperature $T$ when the thermal energy $k_\mathrm{B}T$ becomes comparable to the bending stiffness $B$ of the sheet. However, the crumpling transition has never been convincingly observed in actual experiments, presumably because the critical temperature comes out to be unphysically high \cite{yllanes2017thermal}.   

In this paper, we perform highly sensitive and well-controlled crumpling experiments on extremely thin sheets and report for the first time experimental observation of a crumpling transition whereby a thin sheet, when subjected to a compression larger than a critical threshold, gives up all resistance to further compression and begins to spontaneously evolve toward a more crumpled state. We show that the crumpling transition observed in our experiments is driven by a competition between the surface energy of the sheet, rather than by entropy, and its elastic energy. The transition exhibits classic signatures of a critical phenomenon, e.g. a diverging length scale and power-law scaling near the transition. Our experiments uncover a new and interesting class of elastocapillary phenomena with immense potential for doing controlled origami with extremely thin sheets.


\section*{Capillary crumpling transition}
We illustrate the mechanism underlying this transition through a schematic in Fig. \ref{fig:schematic} A, showing a flat sheet being compressed progressively into a small crumpled ball. The degree of compression is characterized by $\tilde{W}(=W/W_0)$, the ratio of the size of the crumpled sheet ($W$) to the initial sheet size in the flat state ($W_0$). Any process that compresses the sheet (reduces $\tilde{W}$) costs deformation energy ($U_\epsilon$). On the other hand, initial compression does not cause any significant change in the surface area and surface energy ($U_\gamma$) of the sheet. However, below a threshold $\tilde{W}_c$ the sheet begins to make self-contact resulting in a decrease in $U_\gamma$ as the exposed surface area of the sheet is reduced. If the change in surface energy in this regime dominates over the deformation energy cost, the sheet may continue to spontaneously compress into a more crumpled state, giving rise to a crumpling transition at $\tilde{W}=\tilde{W_c}$. 

\section*{Capillary crumpling number}
The stress in a crumpled sheet is localized in singular structures, viz. cones and ridges, and the deformation energy is dominated by that of the stretching ridges \cite{witten2007stress,lobkovsky1995scaling}. We can therefore obtain an estimate of $U_\epsilon$ and $U_\gamma$ in a typical crumpling process by focusing on a single ridge. For a sheet of thickness $h$ and Young's modulus $E$, the area and the deformation energy of a stretching ridge scales with its length $X$ as $X^2(h/X)^{1/3}$, and $B (X/h)^{1/3}$, respectively, where $B\sim Eh^3$\cite{lobkovsky1995scaling}. Therefore, assuming $X\sim W$, we get $U_\epsilon\sim B (W/h)^{1/3}$ and $U_\gamma\sim \gamma W^2(h/W)^{1/3}$. We define a dimensionless \emph{Capillary crumpling number}: 
$Cr=U_\gamma/U_\epsilon\sim\frac{\gamma}{Y}(\frac{W}{h})^{4/3}$, where $Y\sim Eh$ represents the stretching modulus of the sheet. While for $Cr<1$, surface energy is insignificant, for $Cr>1$, beyond a critical compression $\tilde{W}=\tilde{W}_c$, the gain in surface energy outweighs the deformation energy cost and the sheet spontaneously begins to self-crumple. A corresponding phase diagram in the $\tilde{W}-1/Cr$ plane is schematically represented in Fig. \ref{fig:schematic} B. For a large class of macroscopic sheets on which controlled crumpling experiments have been performed \cite{cambou2011three,lahini2017nonmonotonic,shohat2023logarithmic,gottesman2018state,gottesman2015furrows,matan2002crumpling,aharoni2010direct}, we find $Cr \lesssim 1$. On the other hand, we find that for many systems of practical importance in nanotechnology and biology $Cr>1$ (see Table S1). It is difficult to perform controlled crumpling experiments on such systems, and only a few previous works have considered the role of surface energy in crumpling\cite{Liao2021,Nagashima2021}.

\section*{The experiment}
We introduce an extremely sensitive technique to compress thin sheets weakly adhered to solid substrates in a well-controlled manner, and use it to study crumpling in the $Cr>1$ regime. Figure \ref{fig:schematic} C shows a schematic of the experimental method~\cite{Aashna_2024,methods}. Polystyrene sheets cut into circles of initial radius $W_0$ are transferred on hydrogel substrates swollen with water (initial radius $R_0$). The hydrogel substrates are then allowed to shrink as the absorbed water slowly evaporates out, as a result the sheets experiences a small compressive force due to the weak frictional coupling with the substrate. We monitor the radius of the substrate ($R(t)$) and the sheet ($W(t)$) and plot $\tilde{W}$ ($=\frac{W(t)}{W_0}$) as a function of $\tilde{R}$ ($=\frac{R(t)}{R_0}$) (Fig. \ref{fig:schematic} E). The slope of the $\tilde{W}(\tilde{R})$ curve, $\chi=\frac{\mathrm{d}\tilde{W}}{\mathrm{d}\tilde{R}}$ (Fig. \ref{fig:schematic} F), represents the mechanical response of the sheet to the compressive frictional force at the interface. Since the intrinsic friction between a flat hydrogel-I substrate and a flat sheet is very small, we choose a spherical substrate geometry for many of our experiments, where the geometrical incompatibility between the sheet and the substrate allows us to attain sufficiently large friction~\cite{Aashna_2024} to generate enough compression in the sheet.

\section*{Experimental observation of the transition}

\begin{figure}[h]
    \centering
    \includegraphics[scale=0.88]{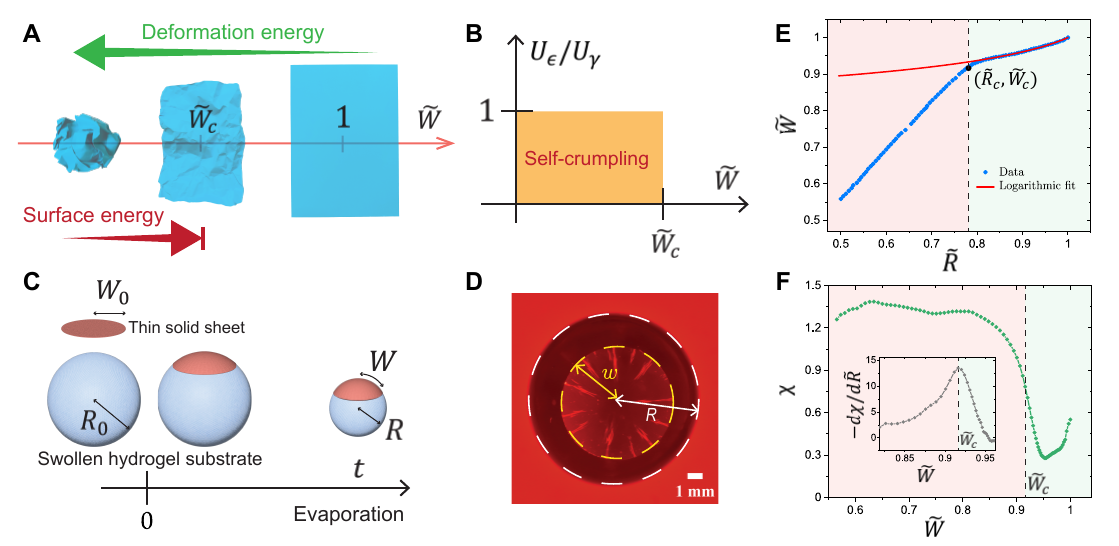}
    \caption{\textbf{Capillary crumpling transition:} (A) A schematic representation of the interplay between deformation and surface energies during crumpling of a thin sheet. (B) If the deformation energy $U_\epsilon$ is less than the surface energy $U_\gamma$, we get a self-crumpling phase for $\tilde{W}<\tilde{W_c}$. (C) The experimental scheme employed for controllably compressing a thin solid sheet. (D) A typical top-view image of the sheet (yellow dashed circle) on the hydrogel substrate (white dashed circle) used to measure $\tilde{W}(t)$ and $\tilde{R}(t)$. (E) Experimental data showing $\tilde{W}(\tilde{R})$. Initially ($\tilde{R}\sim\tilde{W}\sim1$), the sheet compresses slowly: $\tilde{W}=1-a\log(1+b(1-\tilde{R}))$, with fitting parameters $a$ and $b$ (red line). After reaching a threshold point ($\tilde{R}_c, \tilde{W}_c$), the sheet begins to compress much more rapidly. (F) The sheet’s response $\chi=\mathrm{d}\tilde{W}/\mathrm{d}\tilde{R}$ increases rapidly near $\tilde{W}=\tilde{W_c}$. Inset: The quantity $-\mathrm{d}\chi/\mathrm{d}\tilde{R}$ shows a peak at the transition.}
    \label{fig:schematic}
\end{figure}

In Fig. \ref{fig:schematic} E, we start from $\tilde{R}=\tilde{W}=1$ at $t=0$, with a small non-zero slope, which usually becomes smaller with increasing sheet compression consistent with the common experience of crumpling paper, where it becomes harder and harder to compress the paper further. We usually find in this regime $\tilde{W}=1-a\log(1+b(1-\tilde{R}))$ (red line in fig. \ref{fig:schematic} E). As discussed in Supplementary Text, this functional form is consistent with previously observed logarithmic relaxation in crumpling of thin sheets under constant external load and suggests that the compressive frictional force on the sheet remains nearly constant in our experiments. However, as the system reaches a critical point ($\tilde{R}_c$, $\tilde{W}_c$), we observe that $\tilde{W}$ suddenly begins to decrease rapidly with $\tilde{R}$, $\chi$ rises sharply to a value $\chi_\mathrm{s}$ close to $1$ (Fig. \ref{fig:schematic} F), and the second derivative $-\frac{\mathrm{d}^2\tilde{W}}{\mathrm{d}\tilde{R}^2}$ (Fig.\ref{fig:schematic} F inset) goes through a peak. This event marks a transition in the sheet's response to compression, the nature, origin, and significance of which we discuss in the following sections. 

\section*{The transition point}

\begin{figure}[h]
    \centering
    \includegraphics[scale=0.65]{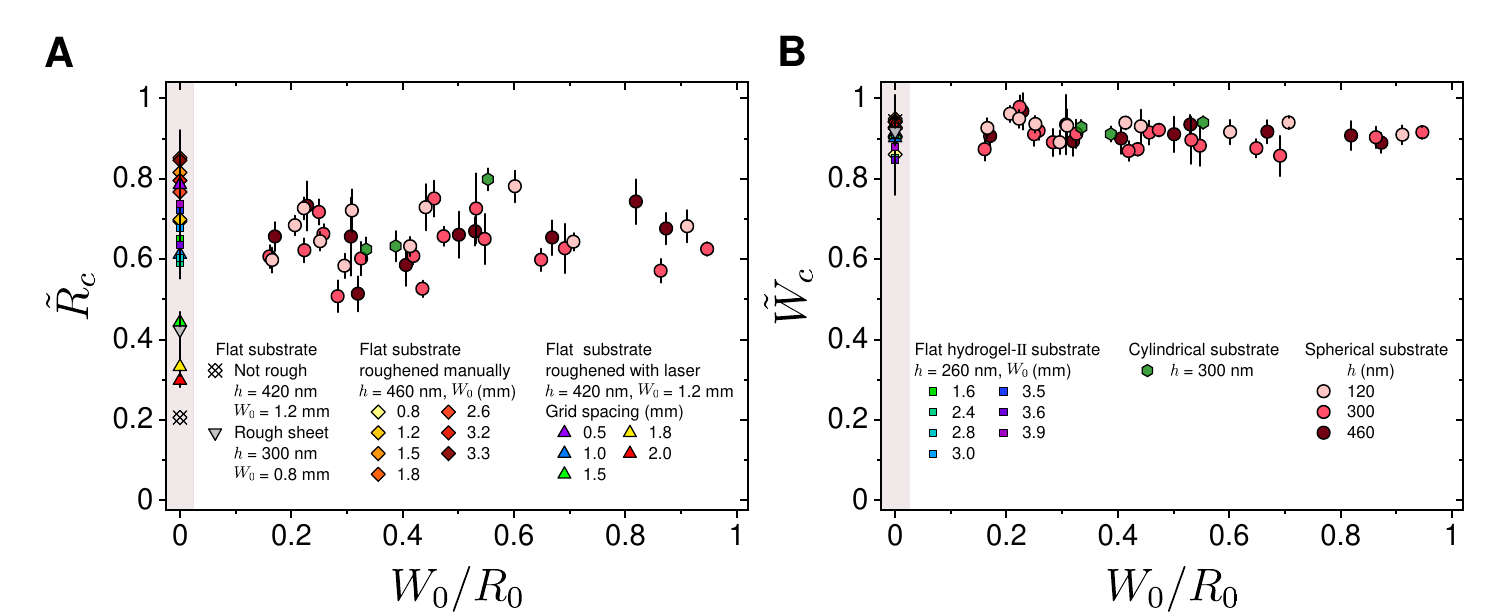}
    \caption{\textbf{Transition point:} (A) $\tilde{R}_c$ and (B) $\tilde{W}_c$ plotted as a function of $W_0/R_0$. $W_0$ is the initial radius of flat sheets, and $R_0$ the initial radius of the spherical and the cylindrical substrates. All data on flat substrate is plotted at $W_0/R_0=0$. The legends describe the various substrate and sheet types on which experiments have been performed. Legends where substrate material is not explicitly mentioned correspond to hydrogel-I. See Supplementary Text for more details.}
    \label{fig:transitionpts}
\end{figure}

We plot in Fig. \ref{fig:transitionpts} A and B the values of  $\tilde{R}_c$ and $\tilde{W}_c$, respectively, as a function of $W_0/R_0$ observed in a large number of experiments performed on sheets with $h$ ranging from $120$ nm to $460$ nm, and $W_0$ ranging from $0.8$ mm to $5.2$ mm, placed on spherical ($R_0\sim 6$ mm) (Movies S1 to S3), cylindrical (Movie S4) and flat substrates (Movie S5 and S6). We have used two different hydrogel materials- commercially available hydrogel spheres (hydrogel-I) and polyacrylamide hydrogel substrates prepared in our lab (hydrogel-II). In the case of flat and cylindrical hydrogel-I substrates either the substrate or the sheet has been additionally roughened to enhance the otherwise small frictional interaction with the sheet~\cite{methods}. We robustly observe the transition in all these different experiments.

The values of $\tilde{R_c}$ and $\tilde{W_c}$ offer the first clue to the mechanism underlying the observed crumpling transition. We notice that while $\tilde{R}_c$ has a large scatter, the transition occurs at a relatively well-defined value of $\tilde{W}_c$ ($=0.91\pm 0.03$) in all the experiments performed under varying conditions. These observations suggest that the transition may not be caused by any change in the sheet-substrate interaction that may arise due to the change in the water concentration in the hydrogel, instead they point towards a generic mechanism related to the geometry of thin sheet deformation during crumpling. A further insight is provided by the value of $\chi_\mathrm{s}$ attained after the transition (Fig. \ref{fig:relaxation} A, see SI for details)--it not only reaches $1$, the limit of complete compliance where the sheet compresses at the same rate as the substrate, in many cases it exceeds $1$, representing, quite remarkably, a tendency of self-compression in the sheet.

\section*{Relaxation experiments}
\begin{figure}[h]
    \centering
    \includegraphics[scale=0.44]{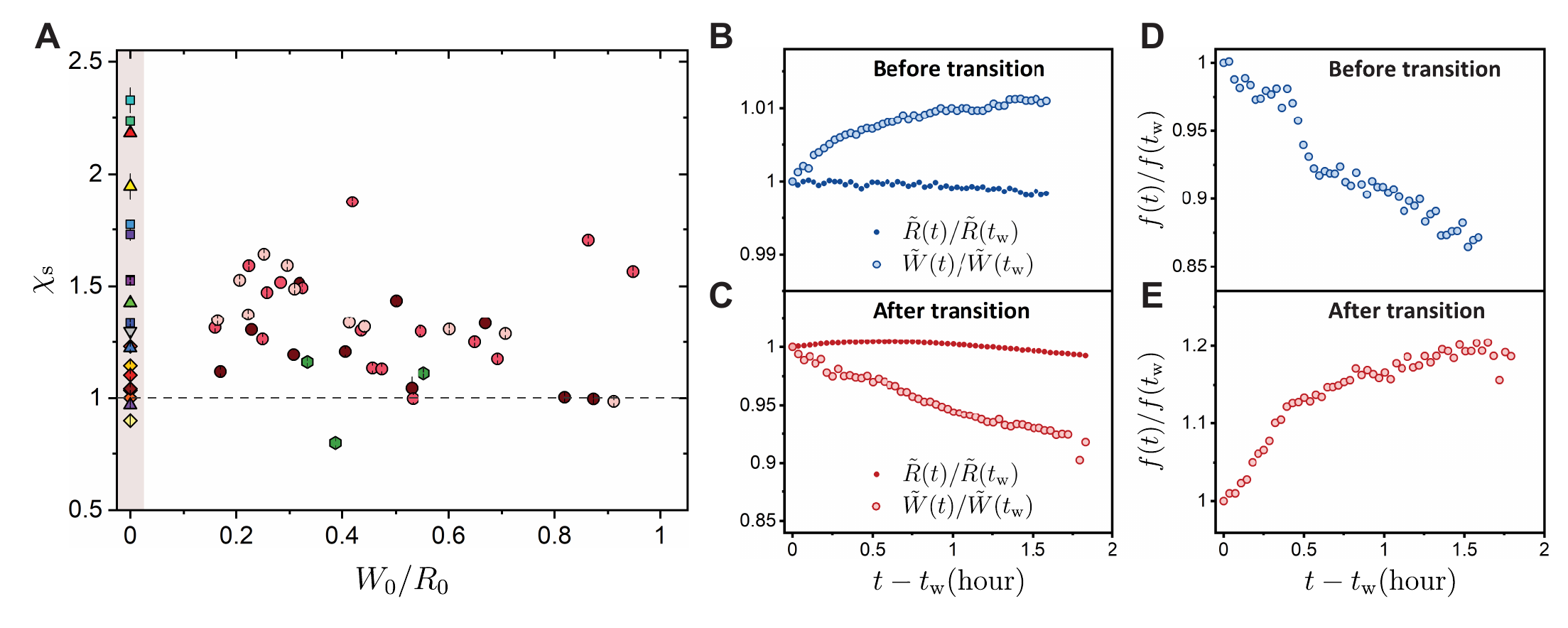}
    \caption{\textbf{Self-crumpling after transition}: (A) Variation of $\chi_\mathrm{s}$, the slope of $\tilde{W}(\tilde{R})$ curve after transition, with $W_0/R_0$. A value of $\chi_s \geq 1$ implies that after the transition the sheet compresses faster than the substrate. The symbols have the same meaning as described in the legend in Fig. 2. The relaxation of $\tilde{W}$ once the experiment is stopped at time $t_\mathrm{w}$ (B) before, and (C) after the transition (large circles). The observed variation in $\tilde{R}$ is also plotted for comparison (small dots) in the respective graphs. The relative change in the fold density $f$ observed during these two time periods are plotted in (D) and (E), respectively. While on stopping the experiment before the transition, the fold density decreases and the sheet relaxes towards an open state, it keeps compressing further on its own with increasing fold density after the transition. }
    \label{fig:relaxation}
\end{figure}

The emergence of self-compression in the sheet for $\tilde{W}<\tilde{W}_c$ is further confirmed by the relaxation experiments, where after running the experiment for a certain time ($t_\mathrm{w}$) we \emph{stop} the experiment by isolating the whole system in a high humidity environment, suppressing evaporation, and therefore shrinking of the substrate. Figure \ref{fig:relaxation} B and C show the relaxation observed in $\tilde{W}(t)/\tilde{W}(t_\mathrm{w})$ before and after the transition, respectively. If we stop \emph{before} the transition, $\tilde{W}$ begins to increase with time--the sheet uncompresses and opens towards its initial undeformed state. However, if we stop \emph{after} the transition, $\tilde{W}$ keeps decreasing--the sheet continues to compress further on its own. The contrasting relaxation behavior before and after the transition clearly shows that it is associated with the emergence of a tendency for self-compression in the sheet.

Our fluorescent imaging technique allows us to visualize the evolution of fold structures, representing the microscopic state of the sheet, during the experiments (Fig.\ref{fig:schematic} D, Supplementary Text). From the fluorescence images, we compute the fold density $f$, as the area of the sheet covered by folds divided by the sheet's total area, and plot $f(t)/f(t_\mathrm{w})$ during the relaxation experiments performed before (Fig. \ref{fig:relaxation} D) and after (Fig. \ref{fig:relaxation} E) the transition. When we stop before the transition, the folds recede and $f(t)/f(t_\mathrm{w})$ becomes smaller with time. However, when we stop after the transition, the folds continue to grow, and $f(t)/f(t_\mathrm{w})$ increases with time.

The contrasting nature of relaxation before and after the transition shows that the transition separates two different phases of the system--before the transition, the sheet prefers a flat configuration and left on its own, it slowly relaxes towards it, however, after the transition, the sheet tends to spontaneously evolve towards a more compressed and crumpled state. This clearly shows that the observed transition is a type of \emph{crumpling transition}. However, it must be noted that our system is athermal, and this transition cannot be driven by entropic forces but, some other mechanism.

\section*{Growth of folds near the transition}
\begin{figure}[h]
    \centering
    \includegraphics[scale=0.5]{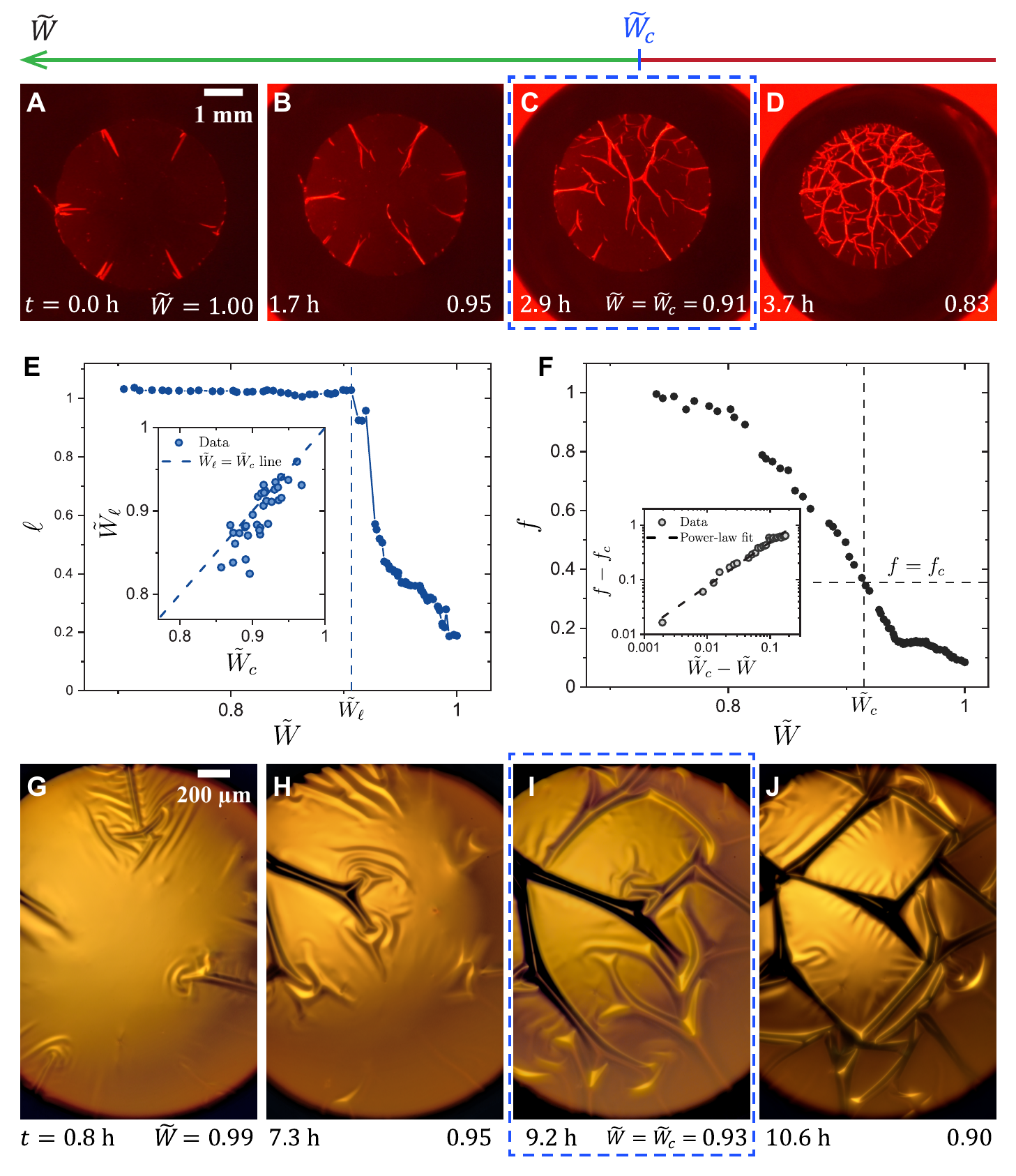}
     \caption{\textbf{Growth of folds near the transition}: (A)-(D) Fluorescence images of the sheet ($W_0 =2.3$ mm, $h =300$ nm) at various times across the transition. Near the transition point (panel C) the folds grow rapidly and a system spanning fold network emerges. (E) The variation of $\ell$ with $\tilde{W}$. Near the transition, $\ell$ increases rapidly and attains the value $1$, at $\tilde{W}=\tilde{W}_\ell$, reflecting the percolation phenomena. Inset: A Comparison of $\tilde{W}_\ell$ and $\tilde{W}_c$ shows that their values lie close to each other. (F) Variation of $f$ with $\tilde{W}$. Inset: A log-log plot of $f-f_c$ versus $\tilde{W}_c-\tilde{W}$ after the transition. The dashed line is a power-law fit $ f-f_c=A(\tilde{W}_c-\tilde{W})^\alpha$ giving an $\alpha=0.81$. (G)-(J) Optical microscope images of the folds and their evolution across the transition, confirm the emergence of fold percolation at the transition. }
    \label{fig:folds}
\end{figure}

A close look at the structural evolution of folds across the transition helps us to understand the underlying mechanism. Figures \ref{fig:folds} A-D (Movies S1 to S3) show a sequence of fluorescence images across the transition. Initially, we observe a few disconnected radial folds near the edge of the sheet which grow as the sheet is compressed. In particular, we notice that just before the transition one of the folds grows rapidly and percolates through the sheet. The same observation is confirmed by microscopy images (Fig.  \ref{fig:folds}G-J, and Movie S7) where we see the deformations directly, though over a smaller region near the center of the sheet.  We track the length of the largest fold $L(t)$ and plot in Fig. \ref{fig:folds} E the evolution of $\ell=L/2w$ with $\tilde{W}$. Here $w$ represents the projected radius of the sheet as opposed to $W$ which is corrected for its curved geometry~\cite{methods}. Near the transition, $\ell$ grows sharply and saturates to a value close to $1$, representing the percolation of fold network through the sheet. 

For each of our experiments, we note down $\tilde{W}_\ell$, the value of $\tilde{W}$ when $\ell$ first reaches the value $1$, and plot it against $\tilde{W}_c$ in the inset of Fig.\ref{fig:folds} E. They are strongly correlated with each other and lie close to the $\tilde{W}_\ell=\tilde{W}_c$ line. It may be noted here that the fluorescence images show fold lengths with a slight time lag compared to the actual fold length as visible under a microscope (Supplementary Text), which may be responsible for the scatter of points slightly below the $\tilde{W}_\ell=\tilde{W}_c$ line.

We also plot in Fig. \ref{fig:folds} F, the evolution of the fold density $f(\tilde{W})$. Near $\tilde{W}=\tilde{W}_c$, we notice that $f$ begins to increase rapidly following a power-law behavior, as evident from the inset of Fig. \ref{fig:folds} F where we plot the change $f-f_c$ with $\tilde{W}_c-\tilde{W}$ on log-log scale; here $f_c=f(\tilde{W}_c)$. The behaviors of both $\ell$ and $f$ are reminiscent of critical phenomena, viz. the existence of a diverging correlation length and power law scaling of certain quantities near the transition.

\section*{Theoretical model}
Motivated by the behavior of folds near the transition, we build a simplified model to explain the transition pathway in our system. For $\tilde{W}<\tilde{W}_\ell$, where we typically observe folds that start from the edge and end inside the sheet, we assume a d-cone-like fold which starts with a semi-circular profile of radius $r_o$ at the sheet edge, and ends inside the sheet in a singular crescent shape (Fig. \ref{fig:folds}) having radius $r_c$. The deformation energy of such a cone-shaped fold scales with the fold length $L$ Supplementary Text) as: $U_{\mathrm{fold}}(L<2W)\sim \frac{\pi B}{r_o^2}\log{(\frac{r_o}{r_c})}L^2$\cite{lobkovsky1995scaling}. As we compress the sheet, the fold becomes longer and both $L$, and $U_{\mathrm{fold}}$ increase. However, as we approach the critical point, $L\to 2W$ and the singular end of the cone is expelled out from the sheet. The fold transitions from being cone-like to being cylinder-like, and its deformation energy can now be estimated as: $U_{\mathrm{fold}}(L=2W)\sim \pi B\frac{2W}{r_o}$. Further, since we have the following hierarchy of length scales in the system: $r_c\ll r_o\ll W$, $r_c\sim h$, and $r_o\sim\ell_{\textrm{bc}}=\sqrt{B/\gamma}$, we find that when a cone-shaped fold percolates the sheet, the singularity at the tip of the d-cone is expelled from the system, and the deformation energy crosses a barrier and decreases significantly (Supplementary Text). In this configuration, the deformation energy cost is outweighed by the gain in surface energy and the sheet begins to spontaneously compress into a more crumpled state.

\section*{Origami with ultra-thin sheets}
\begin{figure}[h]
    \centering
    \includegraphics[scale=0.58]{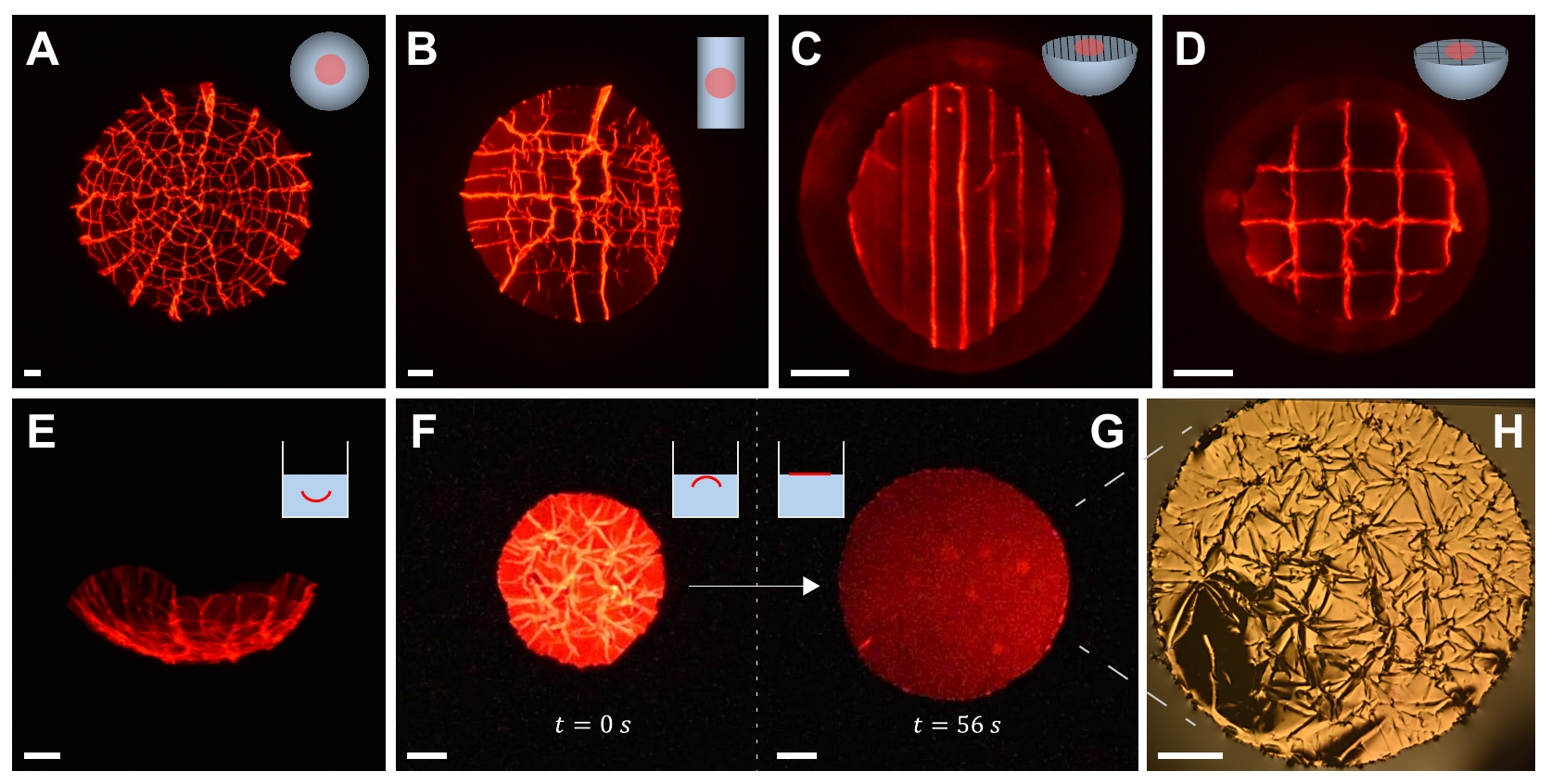}
    \caption{\textbf{A technique for ultra-thin sheet origami:} Fluorescence images of thin flat sheets taken after the crumpling transition ($h\sim 300$ nm) showing the different patterns of folds obtained on (A) spherical substrate (B) cylindrical substrate (C) flat substrate with an array of linear grooves (D) flat substrate with a square grid of grooves. (E) Image of a flat sheet crumpled on a spherical substrate after being released from the substrate inside water showing that the sheet retains its spherical shape. Image of a crumpled sheet inside water (F) just before, and (G) after it is adsorbed onto the water-air interface. (H) Image of the same sheet as in (G) taken under a microscope. Scale bars in all images are $0.5$ mm long. Images (F), (G), and (H) are from the same experiment.}
    
    \label{fig:origami}
\end{figure}

Subsequent to the crumpling transition, the folds in the sheet grow rapidly to form an interconnected network. The symmetry of the fold pattern can be tuned through the geometry of the underlying substrate. This suggests potential for application as a technique to do controlled origami with extremely thin sheets. Figures \ref{fig:origami} A and B show patterns of folds obtained on a spherical and a cylindrical substrate, respectively. On a spherical substrate, in a suitable regime, we obtain a regular array of radial and azimuthal folds, resembling a spider web-like pattern, while on a cylindrical substrate, we obtain a rectangular fold pattern. These observations suggest a phase space where a range of fold patterns is attainable as a function of the two principal curvatures ($\kappa_1, \kappa_2$) of the substrate. The cylinder, and sphere represent only two special cases in the ($\kappa_1, \kappa_2$) plane, corresponding to $\kappa_2=0$, and $\kappa_1=\kappa_2$, respectively, and one expects to obtain other kinds of fold pattern as one explores various parts of the ($\kappa_1,\kappa_2$) plane. One can also look at this system as a physical origami pattern computer--the system finds fold patterns that can transform a flat $2D$ sheet into the $3D$ shape of the substrate. 
 
The range of accessible fold patterns can be further extended by introducing inhomogeneity in the substrate curvature. Figures \ref{fig:origami} C and D show two examples of such an approach where we perform experiments on a flat substrate whose surface has been engraved with an array of equally spaced linear grooves and an array of square grooves, respectively, using a laser cutter. In these cases, we find that the sheets develop fold patterns that closely resemble the pattern of grooves.  

A series of previous works have studied the formation of wrinkle patterns in thin sheets placed on liquid \cite{king2012elastic, tobasco2022exact} and solid \cite{box2023delamination} substrates, however, mostly in the frictionless limit. The above observations open interesting questions regarding fold pattern formation due to friction-induced compression in these systems and its dependence on substrate geometry. 

Moreover, after the crumpling transition, the fold pattern in the sheet and its 3D shape, is trapped in a metastable state. As a result, even when the sheet is now released from the substrate, it continues to retain its shape. In Fig. \ref{fig:origami} E (Movie S8) we show an image of a flat sheet crumpled on a spherical substrate, after it is released from the substrate inside water. The sheet retains its spherical shape. 

We also note that in the controlled crumpling process used here, the deformations are spatially localized and only neighboring regions of the sheet interact with each other. Consequently, a large part of these deformations can be reversed under a suitably large extensile force. Figure \ref{fig:origami} F shows the image of a crumpled sheet, with the fold pattern clearly visible, inside water near the water-air interface. When this sheet adsorbs to the interface, the capillary force pulls open the folds, taking the sheet back to a flat state as shown in Fig. \ref{fig:origami} G (Movie S9). When viewed under a microscope (Fig. \ref{fig:origami} H) we find that the sheet is not completely flat, but has a distinct presence of the crease pattern imprinted on it during the transition, which is expected to modulate the mechanical response of the sheet in a non-trivial manner. 

\section*{Conclusion}

In conclusion, we have discovered a new class of elastocapillary phenomena where a sufficiently thin and therefore soft sheet when compressed beyond a critical threshold, begins to spontaneously compress into a more crumpled state. This capillary crumpling transition is driven by the sheet’s tendency to reduce its surface energy and is observed to occur at a well-defined value $\tilde{W}_c=0.91\pm0.03$. The transition is marked by a rapid increase in the sheet’s response $\chi$ and the fold density $f$. The transition is closely associated with the percolation of a fold network and shows power-law increase in $f$ beyond the threshold, similar to a critical phenomenon. The observations bear an intriguing similarity to another well-studied class of nonequilibrium phase transitions viz. the jamming transition in granular matter\cite{o2003jamming,majmudar2007jamming}--where the coordination number grows sharply near the transition, and which is associated with the percolation of force network and shows power-law increase in the coordination number beyond the transition. There is also a conceptual similarity between the two, as both these transitions have strong geometrical underpinnings and are related to the formation of a critical contact configuration in the respective systems. It is also quite remarkable that the value of $\tilde{W}_c$ observed in our experiments is very close to the square root of the critical packing fraction for the jamming transition in frictionless disks in $2D$\cite{majmudar2007jamming}. A comparison with the jamming transition suggests many interesting future lines of work, e.g. how does this transition extend along the temperature axis? It may also be interesting to verify the properties of the transition in $3D$ for unsupported sheets. 

From an application perspective, the experimental method presented in this paper provides a powerful, yet simple, and scalable technique to do origami with extremely thin sheets, which can be used to controllably imprint patterns of folds, and sculpt flat sheets into desired $3D$ shapes. Combining the immense potential of origami with the rapidly expanding range of functionalities-- chemical, biological, electrical, magnetic, and optical-- that can be fabricated into thin sheet materials can lead to important technological advances finding applications in diverse fields.

\newpage


\section*{Figures}

\newpage


\newpage


\newpage


\newpage


\clearpage 
\bibliography{References_ALL_short} 

\end{document}